\documentstyle[12pt,psfig,aaspp4]{article}
\begin{document}
\def\vg{{\bf g}}
\def\vr{{\bf r}}
\def\grad{\vec\nabla}
\def\fjk{f_{jk}}
\title{ Stability of disk galaxies in the modified dynamics}
\author{ Rafael Brada and Mordehai Milgrom}
\affil{Department of Condensed Matter Physics, Weizmann Institute of
Science, Rehovot 76100 Israel}

\begin{abstract}
General analytic arguments lead us to expect
that in the modified dynamics (MOND)
self-gravitating disks are more stable than their like in
Newtonian dynamics. We study this question numerically, using a
particle-mesh code based on a multi-grid solver for the (nonlinear)
MOND field equation. We start with equilibrium distribution
functions for MOND disk models
having a smoothly truncated, exponential surface-density profiles and
a constant Toomre $Q$ parameter.
We find that, indeed, disks of a given ``temperature'' are locally more
stable in MOND than in Newtonian dynamics. As regards global instability
to bar formation, we find that as the mean acceleration in the disk
is lowered, the stability of the disk is increased as we cross from the
Newtonian to the MOND regime. The degree of stability levels off deep
in the MOND regime, as expected from scaling laws in MOND. For the
disk model we use, this maximum degree of stability 
is similar to the one imparted to a
Newtonian disk by a halo three times as massive at five disk scale
lengths.
\end{abstract}

\clearpage
\section {Introduction}
\label{stability_intro}
\par
Underlying MOND is the assumption that galaxies do not posses a significant
dark halo. As pointed out by \cite{ost_peeb_crit}, 
 a massive dark halo may be an important  stabilizing agent
of galactic disks. It is thus interesting to compare the stability of 
bare disks in MOND to the stability of similar
 Newtonian disks with dark halos.
Such considerations may also provide a MOND explanation (see \cite{mgst}) 
of the revised Freeman law whereby
the distribution of central surface brightnesses of galactic
 disks appears to be cutoff
rather sharply above a certain surface brightness $B_0$,
 (see e.g. \cite{mcgaugh96} for a recent review and references).
 Translating this value of $B_0$
into a mean surface density (for exponential disks)
 one obtains a limiting surface density that
 is nearly
$\Sigma_0 \equiv a_0G^{-1}$, with $a_0$ the acceleration
 constant of MOND.
In MOND, disks with a mean surface density $\Sigma>>\Sigma_0$ have a different
dynamical behavior than those with $\Sigma<\Sigma_0$.
 In particular the former are
Newtonian and thus are beset by the well-known instabilities
 of bare Newtonian disks.
The latter are more stable locally (as shown in \cite{mgst} using
 perturbation theory)
 and, as we shall show in this work, are also more stable globally.
Global added stability is  also supported by preliminary N-body
calculations carried out by \cite{chst}, and by \cite{griv95}.  The Freeman
law, which asserts that the former type of disk is rare, may
 result from this disparity.
\par
Toomre showed that in Newtonian dynamics a disk is stable to
 all local axisymmetric disturbances  at radius R
if the dimensionless quantity
 \begin{equation} Q(R)= \frac{\sigma_R\kappa}{3.36 G \Sigma} > 1,
\label{Q_newt}\end{equation}
 where $\sigma_R$ is the radial velocity dispersion, $\kappa$ is the 
epicycle frequency, and $\Sigma $ is the surface
density. The criterion in MOND is obtained by simply
replacing $G$ by $ G/\mu^{+}(1+L^{+})^{1/2}$, where $\mu^{+}$ is the
value of the MOND interpolating function, 
$\mu $, just above the disk, and $ L = d\ln\mu(x)/d\ln(x)$ (\cite{mgst}). 
Although Toomre's criterion
rests on local analysis, it is found empirically that the 
condition $ Q>1$ everywhere
in the disk is  a necessary and sufficient  condition for
 global axisymmetric stability.
Stellar disks are always stable to local non-axisymmetric disturbances 
(\cite{gold-lynd,jul-toom}).
Numerical simulations have shown that stellar disks
 are subject to global non-axisymmetric
instabilities, especially the bar instability. This result was
 confirmed  analytically
for a few models by linear, normal-mode analysis.
 The majority of rotationally supported,
self-consistent disk models studied to date by numerical simulations
 and analytical global
 analysis
 are violently
 unstable to bar formation.
However, these simulations do not
reveal the mechanism of the instability nor suggest a way to avoid it.
\cite{toom_swing} suggested a mechanism for the bar instability
 based on what he
named swing amplification.

 Even the Newtonian-plus-dark-matter case the stability problem is
anything but resolved.
 So,  we shall not focus our work on testing for absolute stability in MOND.
Instead we perform a comparative study
between the added stability given to the disk by MOND, and that given
by a dark halo.
In particular we shall ask to what extent can MOND replace the halo's contribution
to the stability of disk galaxies.

Even for Newtonian gravity one lacks simple analytic equilibrium solutions
of the collisionless Boltzmann equation for a thin disk. Some of 
the equilibrium
models studied to date are the Isochrone
 by \cite{kalnajs1978}, Kuzmin-Toomre (\cite{sell86,hunt}),
 and the Sawamura disks (\cite{saw}). The extent to which the
 bar instability and others depend on the specific
 properties of these models is unknown.
 The situation for MOND is even more difficult since we have no analogous analytical models. The analytical methods used  in \cite{kaln72} and \cite{kaln77} for  linear, normal-mode analysis
 are very cumbersome and give no physical insight into the nature of the instability.
A simpler way to get the unstable modes is through 
N-body simulations.
We have developed a three-dimensional,
 N-body code and potential solver for the
nonlinear MOND problem, in which the potential is determined from the
equation proposed by \cite{mgbk}.

\section{Description of the MOND potential solver}

\cite{mgbk} have formulated a non-relativistic 
Lagrangian theory for MOND, in which the acceleration
field $\vg$ produced by a mass distribution $\rho$ is derived
 from a potential $\phi$ ($\vg=-\grad\phi$) satisfying
the equation \begin{equation} \label{mond_field_eq}
\grad\cdot[\mu(|\grad\phi|/a_0)
\grad\phi]= 4\pi G\rho \end{equation}
 instead of the usual Poisson equation 
$\grad\cdot \grad\phi=4\pi G\rho$,
where  $\mu(x) \approx x$ for $x\ll 1$, and
$\mu(x)\approx 1$ for $x\gg 1$,
 and $a_0$ is the acceleration
constant of MOND. The form  $ \mu(x)=x/\sqrt{1+x^2}$ has been used
in all rotation curve analyses and we also use it here.  A solution
 to the field equation exists and is
 unique in a domain $D$ in which $\rho(\vr)$ is given and
on the boundary of which $\phi$, or
 $\mu(|\grad\phi|/a_0)
\partial_n\phi$, is specified (\cite{mgfe}). 
In this theory, the usual conservation
laws of momentum, angular momentum, and energy (properly defined) hold,
and, in addition, the center-of-mass acceleration of 
a star or a gas cloud in the field of a galaxy  obeys the basic MOND
 assumption even if its internal accelerations are high.

The nonlinearity of the MOND field eq.(\ref{mond_field_eq})
prevents one from using the standard potential solvers (force
calculators), at least in a straightforward way.
We wrote a multigrid  solver for the finite difference approximation
of the MOND field equation and incorporated it into an N-body code
using the particle-mesh algorithm.
 We give a brief description of the N-body code in Appendix \ref{nbody_prog},
and also describe there some of the tests we have performed to establish its
accuracy, and the setup of initial conditions for the simulation.
We lack analytical potential density pairs for disks in MOND
(apart for that for the Kuzmin disk), not
to mention self-consistent stationary models. We have thus
developed a numerical scheme for generating self-consistent stationary 
disk models with specified potential, surface density, and radial
velocity dispersion. This scheme is described in appendix
 \ref{model_construction}.
\par
\begin{figure}
\centerline{\psfig{figure=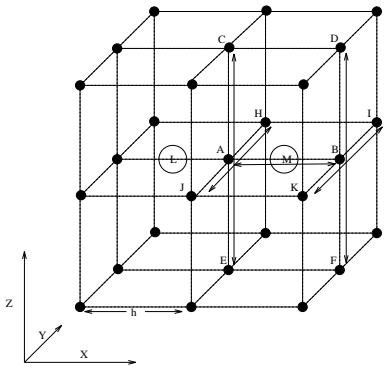,width=5cm}}
\caption{ The discretization of the MOND equation. }
\label{fig1}
\end{figure}

Because the potential solver is novel we describe it here briefly.
The discretization scheme used is depicted in Figure~\ref{fig1}.

It uses central differencing between neighboring grid points
to approximate the divergence and the components of  $\grad\phi$
 appearing in eq.(\ref{mond_field_eq}).
Only for some of the components of $\grad \phi $ appearing in the
  argument of the function $\mu$ do we
use central differencing between grid points that are two grid spacing 
apart.
The $\partial_x$ part of the divergence in
 eq.(\ref{mond_field_eq}) at point A is  approximated by
$ [S(M)-S(L)]/h \;$, where
 $ \;S=\mu(|\grad\phi |/a_0)
 \partial_x \phi$.  
  $\;\partial_x\phi(M) $ is 
 approximated by $ [\phi (B)-\phi (A)]/h $, 
$ \; \partial_y\phi(M) $   by 
 $[\phi (I)+\phi (H)-\phi (J)-\phi (K)]/(4h)$,
and $\partial_z\phi(M) $  by
 $[\phi (C)+\phi (D)-\phi (E)-\phi (F)]/(4h)$.
A similar calculation is done for point L and for the remaining 
parts of the divergence.
 This is a stable
second order discretization, which, importantly, is  flux conserving.
The  MOND equation, like Poisson's,
 can be transformed using Gauss's theorem into a flux equation
\begin{equation} \int_{\partial D} \mu\frac{\partial \phi}{\partial n}
 dS =\;4\pi G\; \int_{D}  \rho d^3x,
\label{eq:mgauss} \end{equation} where $D$ is any domain where the MOND 
equation is satisfied,
$\partial D $ is the boundary of $D$ and  $\partial \phi /\partial n $
 is the normal derivative of $\phi$. The flux leaving
 a cell through one of its sides
should be equal to the flux entering its neighboring cell; flux
 conservation means that the two will have the same approximation
in the discrete equation.

 We use the multigrid techniques 
 developed by Brandt and collaborators
 (\cite{achi1977,achi-mgguide,achi-lattice91,achi-local-ref1987}),  
which is extremely efficient in solving elliptic, partial
differential equations. We use the so-called full-multigrid
 algorithm together with the
full-approximation scheme,
and use Gauss-Seidel relaxation
for solving the system of nonlinear equations produced by the
 discretization. Instead of solving directly
for the new value of the unknown at the current grid point we
 carry out a single iteration of the
Newton-Raphson method for finding the root of a nonlinear equation,
 where  the derivative of the left-hand
side of the equation with respect to a change in the unknown is
 calculated numerically. 
For solving the standard Poisson equation we use Gauss-Seidel relaxation with
 red-black (RB) ordering, which  
has two important properties: first,  the smoothing rate for the
 usual seven-point Laplacian is the best;
second, the ``red'' and the ``black''  points are independent
 and can be relaxed
 simultaneously. This last property is very useful in writing a code that
 is highly vectorizable and parallelizable.
In order to maintain this property of independence in the case of
 the more complicated MOND equation
we use  a generalization of RB ordering  using eight colors instead of two.
\par
The MOND potential solver was tested extensively against cases for
which exact results are known. These include a. the complete potential 
field of a Kuzmin disk (\cite{bra95}); b. the (deep) MOND, two-body force
for arbitrary masses, and the N-body MOND force in certain
symmetric configurations (\cite{mgMV}); c. a general relation that exists
between the total mass and the root-mean-square velocity for disks
 in the deep MOND regime, first discovered by our numerical calculations,
 and then proven exactly (\cite{mgMV}).


\section{Models and results}
\label{sec_models}
\begin{figure}
\centerline{\psfig{figure=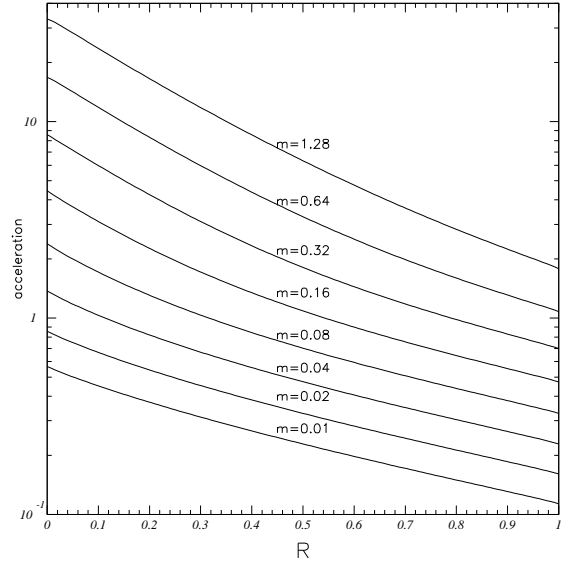,width=8cm}}
\caption{\protect\small The magnitude of the acceleration
 just above the disk (in units of $ a_0$)
 as a function of R (in units of $R_{cut}$) for the different mass models.}
\label{fig2}
\end{figure}

As stated above, we concentrate on a comparative study between
 the stabilizing effects of MOND and those of dark
matter halos. We have used models that have a smoothly truncated,
 exponential surface
density. The disk extends out to radius $R_{cut}=1$ (chosen as our unit of 
length), with a  scale
length of 0.2 in these units. The surface density is of the form
\begin{equation}
\Sigma(R)=\Sigma_0 \exp(-R/0.2) (1-R^{4}), \;\;\;   R \leq R_{cut}
\label{eq_tr_ex}
\end{equation}
The smooth truncation of the disk is used in order to avoid the edge
 instabilities discussed by
\cite{toom_swing}, which result from a sharp drop in the
surface density.
We work in units where $G=1$, $a_0=1$, and the mass  is given in units of 
$a_0 R^{2}_{cut}/G$. We have constructed  a series of  models with
 a total mass of
$ 0.005,0.01,0.02,0.04,0.08,0.16,0.32,0.64,1.28 $. The disk with
 the lowest mass is fully in the
MOND regime ($ a < a_0$),
 while the disk with a mass of 1.28 is
Newtonian almost all the way to its outer edge.
 The magnitude of the  total acceleration 
just above the surface of the disk as a function of radius for the
different  mass models is shown in Figure \ref{fig2}. (This differs from the
mid-plane acceleration, which enters the rotation curve, because of the 
perpendicular component that appears just above the disk.)

A self-consistent model for a given mass 
distribution is also characterized by its ``temperature'':
the fraction of the total kinetic
energy that is in random motion.
A convenient parameter for measuring this is the famous $t=T/|W|$
parameter, where $T= 2^{-1}\int \rho \overline{v_{\theta}}^2 d^3x$
 is the rotational
kinetic energy and $|W|$ is the absolute value of the
 self-gravitational energy, which by the 
virial relation is equal twice the total kinetic energy
 (rotational plus random)  of the stationary system.
In MOND, we replace the self-gravitational energy, in the definition of 
t, with twice the rotational kinetic
energy of a cold system (where all the particles
 are on circular orbits).
The maximum value of $t$ is 0.5 (realized for a cold system).
 The lower the value of $t$,
 the greater the part of the kinetic energy in random motion.
Motivated by the analytical results for the Maclaurin and for the Kalnajs
 disks, and by their own numerical simulations,  \cite{ost_peeb_crit}
suggested the approximate empirical stability criterion
against bar formation $t < 0.14 $. Although the physics of the bar instability is only indirectly
related to $T/|W|$, numerical studies have shown that  this
Ostriker-Peebles  criterion provides a surprisingly useful empirical 
guide for identifying systems that are likely to be unstable.
\par
\begin{figure}
\centerline{ \psfig{figure=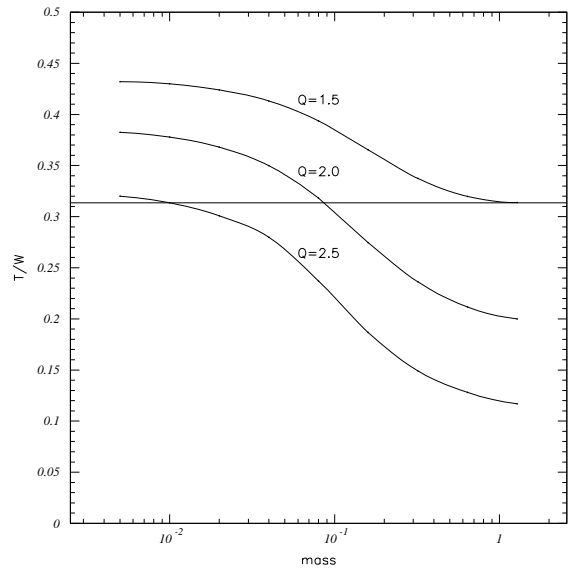,width=8cm}}
\caption{\protect\small~ $T/|W|$ as a function of the disk's mass for
 MOND models with constant $Q=1.5,~2.0,~2.5$. The horizontal
 line at $T/|W|=0.31$, marks the value used in our models}
\label{fig3}
\end{figure}

As a preliminary step in our comparative 
study we generated three self-consistent models for
 each of the total disk masses listed above.
The models have a radius-independent value of the MOND 
  stability parameter [see below eq.(\ref{Q_newt})], with $Q=1.5,2.0,2.5$.
The calculated runs of $t$ values for these models are
displayed in Figure \ref{fig3}.

\par
We then constructed MOND
 equilibrium models for comparative, N-body simulations.
These were taken as
 smoothly truncated, exponential disks with
$t=T/|W|$ fixed at $0.31$, and a value of Q that is $R$
independent. These models fall on the horizontal line in
Figure \ref{fig3}. While the potential field is computed on a 3-dimensional
Cartesian grid, disk particles are, at all times, confined to the mid-plane.
\par
\begin{figure}
\centerline{\psfig{figure=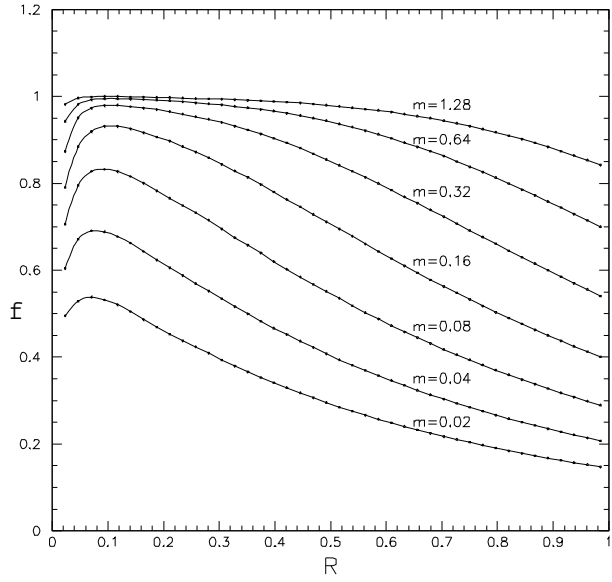,width=8cm}}
\caption{\protect\small The relative contribution of the disk to the total radial
force as a function of radius for the Newtonian-plus-dark-matter cases}
\label{fig4}
\end{figure}

Each model was run once using MOND,
and once using Newtonian gravity.
Because $\Sigma$ and the potential in the plane are the same,
 the Newtonian disk is supplemented
with an inert spherical halo that gives, together with the disk,
 a Newtonian potential that equals
 the MOND potential of the disk alone in the plane.
(The Newtonian disks have the same
 distribution functions as their respective MOND counterparts, but their
 Newtonian $Q$ values are higher, and $r$-dependent, because they do not 
include the $\mu(1+L)^{1/2}$ factor that appears in the MOND expression for 
$Q$.)  
 The lower the total mass of the
disk is the stronger is the departure from Newtonian gravity,
 resulting in an increase in 
the relative contribution of the halo. In Figure
\ref{fig4} are shown the relative contributions of the disk to the total radial
force as a function of radius for the Newtonian-plus-dark-matter cases.

\begin{figure}
\centerline{\psfig{figure=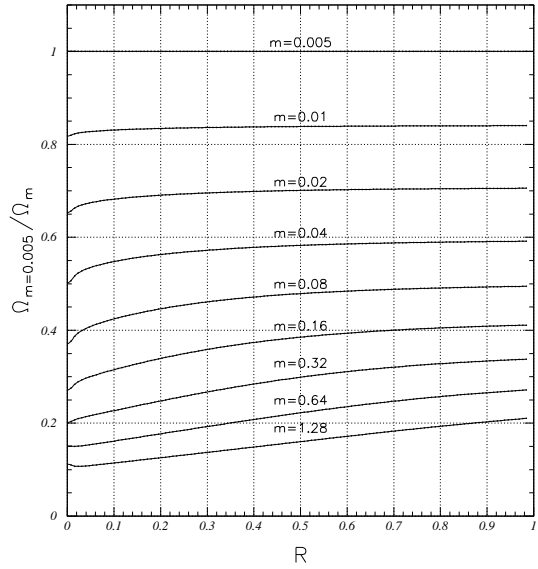,width=8cm}}
\caption{\protect\small The ratio of the angular frequency of the $m=0.005$ mass model
to the angular frequencies of all the mass models as a function of radius}
\label{fig5}
\end{figure}

In order  to make a quantitative comparison between the growth rates of the unstable
modes of the different mass models we scale the time step in the simulation
in proportion to a natural dynamical time of the model. In Figure \ref{fig5}
are plotted, for each model,
 the ratio of the angular frequency of the $M=0.005$ mass model to
to that of the specific model.
 As can be seen from the graph this ratio depends 
somewhat on R. We have chosen to scale the time step in
 proportion to the orbital time
 at $R=0.5$ (scaling 
the time step in proportion  to orbital time at $R=0.25$
does not change  the results qualitatively).

\begin{figure}
\centerline{\psfig{figure=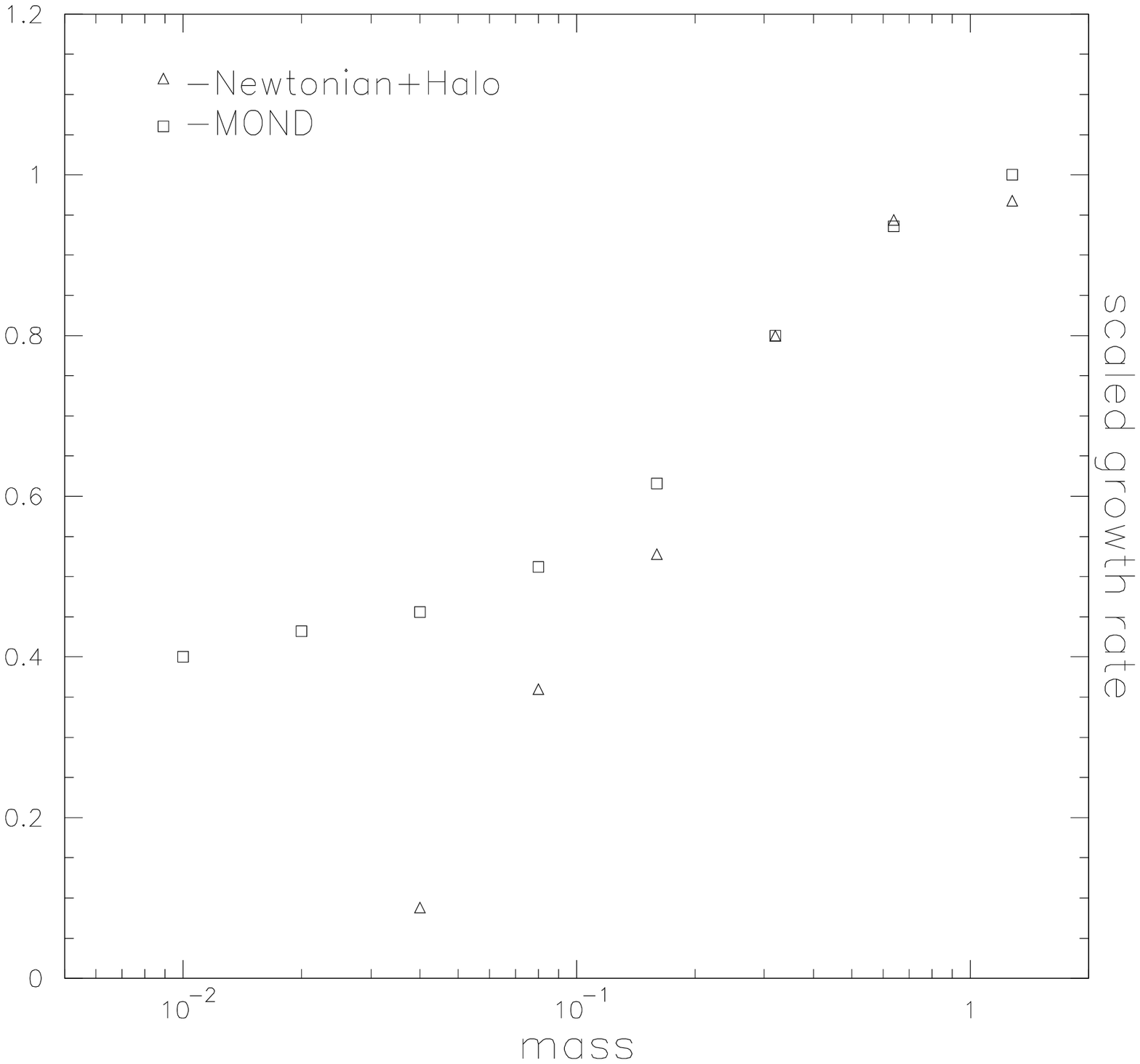,width=8cm}}
\caption{\protect\small The growth rate, in  units of the dynamical
 frequency, for the m=2 mode as a function
of the total mass of the disk. $ \Box$ MOND, $\triangle$ Newtonian + Halo. }
\label{fig6}
\end{figure}

The development of the instability is traced in the time dependence
of  the fraction of the disk's mass in the m=2 Fourier component
of the surface density. 
This turns out to have a period of  exponential growth.
We take the exponential growth rate as a measure of the 
instability's strength. 
In Figure \ref{fig6} we plot the growth rates as functions
  of mass for both the MOND
 and the Newtonian-plus-DM models. These are also given in Table
\ref{table1} together with the Q value and the halo mass (for the 
Newtonian counterparts) of the
different models. The growth rates are calculated using the scaled
time units i.e. the real growth rate equals the growth rate that
 appears in the graph
times $\Omega_{m}(R=0.5)/\Omega_{0.005}(R=0.5)$ as was described previously.

\section{Conclusions}
From the results presented in Figure \ref{fig3}
we see that exponential disks having a given fraction 
of their kinetic energy in  random motion, and a constant $Q$ profile 
are locally more stable in MOND than in Newtonian dynamics,
as reflected in the fact that for the 
same value for $T/|W|$ the MOND disks have a higher  $Q$ value.
This  is in agreement with the general result (\cite{mgst}) 
regarding the local stability of disks in MOND.
One can see that the change in the dynamics occurs when one
 crosses over from the
Newtonian regime to the MOND regime i.e. when the
 acceleration in the disk become
of the order of $a_0$. The added degree of stability is limited (the
change in $Q$ saturates deep in the MOND regime). 
This stems from the fact that at both the Newtonian limit and the deep MOND
limit the equations governing the evolution of the system
 obey simple (but different) scaling laws.
 The basic physical mechanism 
 behind the added stability is
the relatively weaker response in the potential to a given
 perturbation in the surface density
when one is in the MOND regime. Roughly speaking 
in MOND $a^2 \propto \rho$ and therefore
$\delta a/a =\delta \rho/2\rho$, while in the Newtonian
case $\delta a/a=\delta \rho/\rho$, and we see that  approximately
a factor of two is gained in stability.
\par
From the results presented in Figure \ref{fig6} and 
 Table \ref{table1} we see that the global stability of the disk
behaves in a way similar to the local stability.
 As one moves from the Newtonian 
regime to the MOND regime the growth rate  of the $m=2$ mode
 (in  dynamical-frequency units) 
decreases. At first (down to $M\sim 0.2$) the effect of MOND is similar
to that of the added inertial halo. Below that the degree of stability
continues to increase, but not as fast as that of the Newtonian
 disk-plus-halo; and it saturates
 in the deep MOND limit.
In contrast, the Newtonian disk-in-halo becomes increasingly stable in the
limit. 
The saturated global stability given the disk by MOND,
 is similar to that
given a Newtonian disk  by an inert halo with a mass that
 is 2-3 times the mass of the disk  up to
$R=5$ scale lengths. These results support the  idea that pure
 MOND disks with high surface 
densities are less stable than those with a lower surface density both
globally, and locally. This provides a possible  explanation of the
Freeman law as discussed in the introduction.
It must, however, be appreciated that we cannot be sure that actual LSB
 galaxies are more stable than HSB ones because we do not 
know that they all have similar $T/W$ values, as used in our comparison.
\par
Our aim in the paper has been to compare the stability properties in MOND of 
disks with different accelerations. In this, the Newtonian models have
served as references so that the added MOND stability could be described in 
terms of an added inert halo.
 But, what is the significance of the comparison  
between the MOND and Newtonian disks as regards true galaxies?
The disk-plus-inert-halo models we use are not what MOND predicts for a galaxy.
If MOND is correct than in the low acceleration limit there should also
appear to be much disk dark matter. In the present paper we have ignored
the structure and motion in the z direction, perpendicular to the disk.
A Newtonian model that will give the same 3-dimensional
disk distribution function as a MOND pure disk, would have much disk dark
 matter that is not inert, but responds to disk perturbations.
Put differently, MOND predicts that the dynamically determined
surface density of LSB galaxies will be much higher than the observed
surface density. When this surface density is used in calculating the
Newtonian $Q$ value, as it should, a much lower $Q$ value
will result in general (\cite{mgst}).
Inasmuch as we neglect the z-structure and take the disk as infinitely thin
the exact value of the Newtonian surface density that gives the 
same potential field as MOND is $\Sigma_N=\Sigma/\mu^+$, where $\mu^+$ is 
the value of $\mu$  for the local acceleration just above the disk
(because at every point on the disk, and at all times during its evolution, 
we have  $\mu^+\partial_n\phi=2\pi G\Sigma$, while 
$2\pi G\Sigma_N=\partial_n\phi$).
 The net result is that even in the 
deep-MOND limit MOND disks are expected to be somewhat more stable than 
the Newtonian disks that have the same distribution function (and which 
thus have the same r- and z-structure) because of the $(1+L)^{1/2}$ factor.
We do not include here simulations for such Newtonian models.

\appendix
\section{Model construction}
\label{model_construction}
 The problem of finding a distribution function (DF) $f(E,L_z) $ 
 can be made well
 posed for numerical solution by formulating it
 as a constrained optimization 
 problem (see \cite{bt}). One wants  a DF that
 satisfies the following physical requirements as  constraints
 \begin{eqnarray}  
 & f  \geq  0, \nonumber \\ 
 & \int f d {\bf v}  =  \Sigma(r), \\
 & \int f v_r^2 d {\bf v}/ \int f d {\bf v}  =  \sigma_{vr}^2(r),  \nonumber 
 \end{eqnarray}
 and, as an additional auxiliary constraint that will assure uniqueness,
 maximizes a certain functional
 of $f$ such as the Boltzmann entropy or some measure of smoothness.
 A very similar approach is to minimize a single functional
 of $f$ that is the sum of  the errors in the surface density, 
 the radial velocity dispersion, and the entropy,
 subject to the constrain that
 $f\geq 0$. We have used the latter approach.
 We take the disk to be a finite disk  whose surface density vanishes
 for $ r \geq 1$.
 The three input functions: the potential, the surface
 density, and the desired radial velocity dispersion, are
 represented by one dimensional arrays $\phi_i$,
 $\Sigma_i$, and $ \sigma_{vr_i}$, respectively, at
the equidistant grid point $r_i=i/N$ ($0\le i \le N$).
\par
 Using the variables 
\begin{equation}
 X=\frac{L_z^2}{2}-E+\phi(1), ~~~Y=\frac{L_z^2}{2}, \end{equation}
and given a distribution function $f$,
  the surface density and radial velocity dispersion runs  take the forms:
 \begin{eqnarray}
 \Sigma ^\prime (r) & = & \int \int f(X,Y) W(X,Y,r) dX dY,  \label{xysurf} \\
 \sigma^{\prime \, 2}_{vr}  (r) &  =  & \Sigma'^{-1}(r)
\int \int f(X,Y) W(X,Y,r)  Z(X,Y,r) dX dY, \label{xysigvr}  \\
 W(X,Y,r) &  = &  Y^{-1/2}\{ r^2 [ \phi (1)- \phi (r) -X]
 - (1-r^2) Y \} ^{-1/2} \nonumber \\
 Z(X,Y,r) &  = &  v^2_r(r) = 2  [ Y (1-r^{-2} )- X + \phi (1)
 -\phi(r)] \nonumber \\
 0 \leq &  X  & \leq \phi(1) - \phi(r) \nonumber \\
 0 \leq &  Y &  \leq \frac{r^2}{1-r^2} [ \phi(1) - \phi(r) -X]. \nonumber 
 \end{eqnarray}
Note that we work here with the choice $R_{cut}=1$; more generally we would 
have to replace $L_z$ in the above expressions by $L_z/R_{cut}$.
\par
\begin{figure}[htp]
\centerline{\psfig{figure=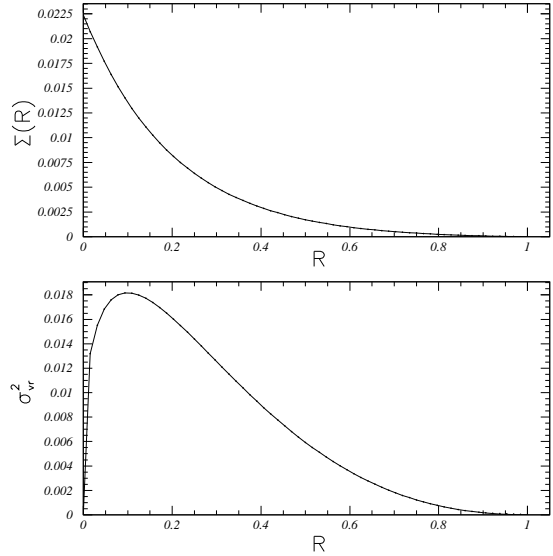,width=8cm}}
\caption{\protect\small The surface density  and the  square of the radial
velocity dispersion for the $m=0.005$,
$Q=2.55$, MOND, truncated exponential disk. The solid line
 represents the model
and the dots are the numerical values calculated from the 
distribution function.} 
\label{fig7}
\end{figure}

 Before discretizing the equations we make a change of variables from $X, Y$ to
 $ r_{min}, r_{max} $, which denote, respectively, the pericenter,
 and apocenter of an orbit with given energy and angular momentum.
 The transformation from $r_{min},r_{max} $ coordinates to the
 $X,Y$ coordinates
 can be obtained by solving the two equations:
 \begin{eqnarray}
 Y & = & \frac{r_{min}^2}{1-r_{min}^2} [\phi(1)- \phi(r_{min}) - X], \\
 Y & = & \frac{r_{max}^2}{1-r_{max}^2} [\phi(1)- \phi(r_{max}) - X]. 
 \end{eqnarray}
 From the expressions $X(r_{min},r_{max})$ and $Y(r_{min},r_{max})$ we calculate
 numerically the Jacobian $\frac{ \partial (X,Y)}{ \partial (r_{min},r_{max})}$
 and rewrite the integrals (\ref{xysurf}, \ref{xysigvr}) using the new coordinates
 where now the limits of integration are:
 \begin{eqnarray}
 0 & \leq & r_{min} \leq r,   \\
 r & \leq & r_{max} \leq 1.  
 \end{eqnarray}
 We also replace $\Sigma'(r)$ in equation~(\ref{xysigvr}) by the  desired
 surface density, $ \Sigma(r) $, since the two become identical when a solution
 is found.
 We discretize the DF  on a Cartesian grid where 
 \begin{eqnarray}
 \fjk = f(r_{max_j},r_{min_k}) \\
 r_{max_j} = (j-1) /N \nonumber \\
 r_{min_k} = (k-1) /N \nonumber 
 \end {eqnarray}
 and the value of f in between grid points is defined using
 bilinear interpolation.
Since interpolation is linear in $\fjk$ and the integrals
 in eqs.~(\ref{xysurf},\ref{xysigvr}),
after the replacement of $\Sigma '$, are also  linear 
in $\fjk$, we can obtain
expressions of the form:
\begin{eqnarray}
\Sigma_i^\prime & = & \sum_{j,k} A_{ijk} \: \fjk,  \\
\sigma_{vr_i}^{\prime \, 2} & = & \sum_{j,k} B_{ijk} \: \fjk/\Sigma_i, 
\end{eqnarray}
where $A_{ijk}$ and $B_{ijk}$ are calculated by numerically
 integrating \\
 $\frac{ \partial (X,Y)} { \partial (r_{min},r_{max})}   W(X,Y,r)$ and
 $\frac{ \partial (X,Y)}{ \partial (r_{min},r_{max})} W(X,Y,r) Z(X,Y,r) $ 
respectively,  over the relevant volume for each grid point.
We are left with the discrete problem of minimizing the expression
 \begin{equation}
 a_1 \sum_{i=1}^{N} \left (\Sigma_i - \Sigma_i^\prime \right )^2 +
 a_2  \sum_{i=1}^{N}
 \left (\sigma_{vr_i}^2 -\sigma_{vr_i}^{\prime \, 2}\right )^2 + a_3 S[f]
\label{terr}
\end{equation}
with respect to the variables $\fjk$ given $ \Sigma_i $ and $\sigma^2_{vr_i}.$
 The functionals of $f$ that we have used
are the Boltzmann entropy which is defined as  $ S=-\int f
 \log f d {\bf x} d {\bf v}$,
or a measure of smoothness which we took as the $L^2$ norm
 of the gradient of f in
the $r_{min}$, $ r_{max} $ coordinates.
After discretization we are left with an expression of the form 
\begin{equation}
S=\sum_{j,k} C_{jk} \: \fjk \log(\fjk)
\end{equation}
for the Boltzmann entropy, and an expression of the form
\begin{equation}
S=\sum_{j,k} D_{jk} [\left(\fjk-f_{j+1~k} \right) ^2 +
 \left(\fjk-f_{j~k+1} \right)
^2]
\end{equation}
for the smoothness functional.
We minimize (\ref{terr}) using an iterative scheme where at each
 step we make a Gauss-Seidel
relaxation sweep, we sweep over the grid, and at each grid point
 we set a new value for $\fjk$
in such a way that it will minimize eq.~(\ref{terr})
 using a quadratic approximation obtained
from the first and second derivatives of eq.~(\ref{terr}) with
 respect to $\fjk$.
After each relaxation sweep we decrease the weight $a_3$ by
 multiplying it by a number 
smaller than one. In this way we let $a_3$ tend towards zero
 as the calculation progresses.
In Figure \ref{fig7} we plot as solid curves the input
 constraints of the surface density
and the square of the radial velocity dispersion and as points
 the values calculated
from a numerical solution found for the distribution function.
 The galaxy model is a smoothly
truncated exponential disk having a total mass of 0.005, a constant $Q=2.55$,
and obeys MOND. These models are described in section~\ref{sec_models}
As can be clearly seen from the graph  the relaxation converges
 to an accurate solution.

\begin{figure}[htp]
\centerline{\psfig{figure=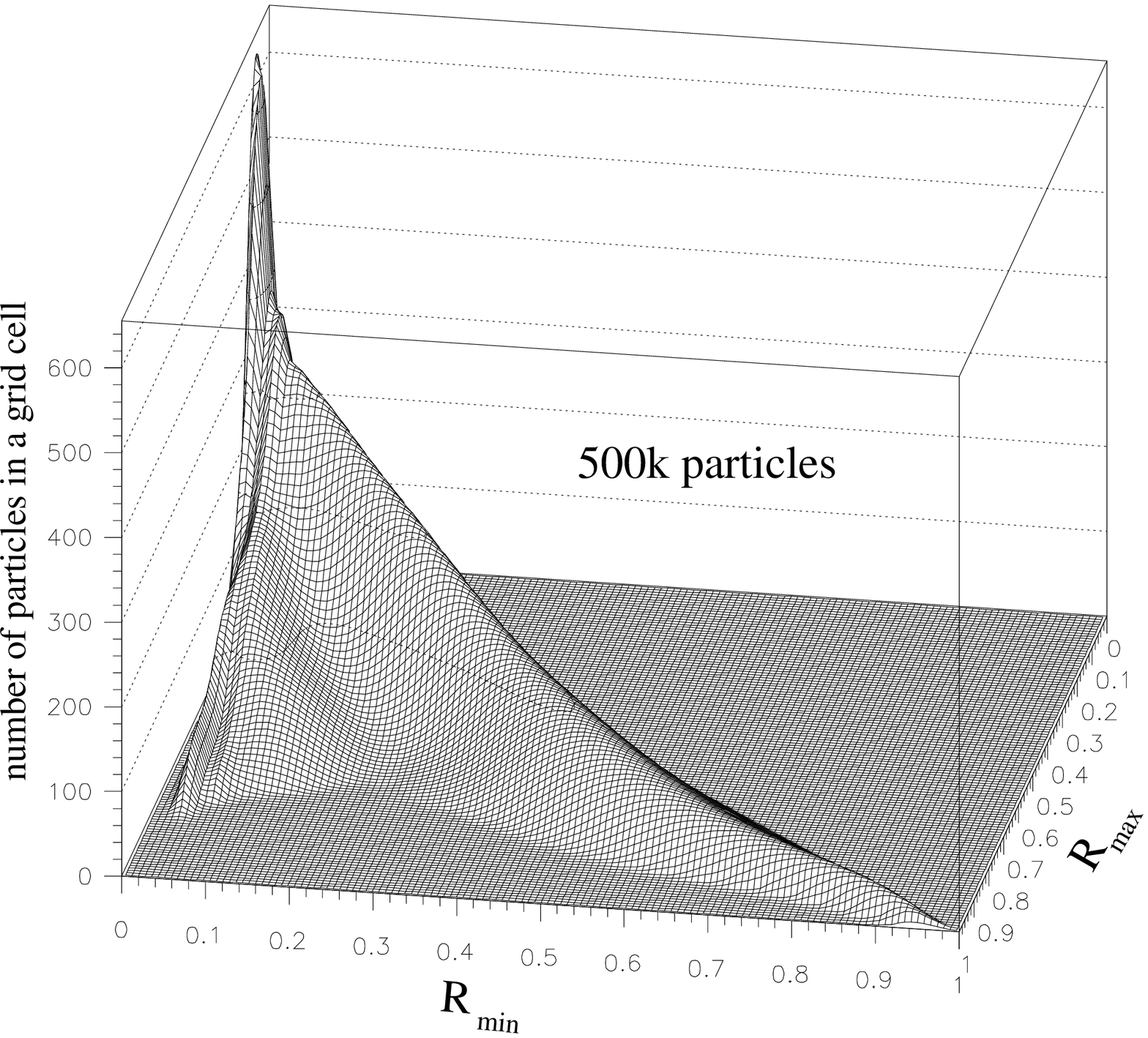,width=8cm}}
\caption{\protect\small The distribution of particles in phase space
for a Q=2.55, MOND, truncated exponential
disk model having a total of 500,000 particles.}
\label{fig8}
\end{figure}

In Figure \ref{fig8} we plot the distribution in phase
 space using $r_{min}-r_{max}$
coordinates of 500,000 particles according to a distribution
 function found for the model.
In Figure \ref{fig8} we plot the distribution in phase
 space using $r_{min}-r_{max}$
coordinates of 500,000 particles according to a distribution
 function found for the model.
\section{An N-body simulation and initial conditions}
\label{nbody_prog}
The nonlinearity of gravity in MOND prevents one from using
  most standard  potential solvers, at least in a straightforward manner.
 Since  we have written a multigrid 
potential solver it is a natural choice to use  the particle-mesh
algorithm, as described, e.g. in \cite{hoea}, for N-body
simulations.
 At each time step  the density is interpolated from the particles to
the grid; then we solve for the potential
on the grid and interpolate
the forces computed on the grid to the particle's location
 in order to integrate its equations of 
motion. We use the cloud in a cell (CIC) charge assignment
 and multi-linear interpolation for the force
calculation at the particle's location; this algorithm is relatively fast.
 The same program can perform a simulation using MOND or Newtonian dynamics.
\par
The potential solvers and the N-body code were extensively
 tested using Newtonian dynamics
and MOND. Once the potential solver has been tested as described at
the end of section 2 and found
 accurate there is no 
difference between Newtonian dynamics and MOND in the rest of the N-body code. 
 The N-body code was tested
by running stable King models, both Newtonian and MOND models,
 and observing the stationarity of the different quantities
such as the size of the system, average velocities, total angular
 momentum, linear momentum, energy, etc.
\cite{kalnajs1978} reported the eigenfrequencies of the dominant bisymmetric
eigenmodes of the isochrone/$m_k$ models. \cite{earn_sell_1995}
 used Kalnajs's distribution functions 
for the isochrone/12,9 models  to compare the results they got
 from their expansion code
to the  analytic results of Kalnajs. We have run a simulation
 using our code, Newtonian dynamics,   
and their initial conditions. We then  performed the same fit as they did
for  the pattern speeds and growth rates of the unstable modes. The fit
between the numerical results and the analytical results were  
 good (about 10-15\% accuracy).
\par
The importance of a careful initial setup for an equilibrium model
 is well documented
 (\cite{sell83,sell_rev87}).
There are two separate aspects to this: suppression of particle noise
 and choosing
coordinates from the desired distribution function.
Initial positions picked randomly produce
 shot-noise density fluctuations
 on all scales. For initial, near-equilibrium models the initial
 behavior is dominated by the collective response to the 
artificial noise, and this can
mask the dominant modes of the continuous system (\cite{sell83}), in which
we are interested.
 Such initial noise can be
 suppressed by arranging the particles regularly.
This results in a discrete noise spectrum with
 large amplitudes 
at the wavelength of the particle spacing, which must be suppressed
 during the force determination.
This would give a particle distribution that behaves as a smooth fluid.
To this end, for our polar grid we place particles
 on rings spaced in radius
 according to the required surface density law.
 The number of particles on each ring must be related
to the number of azimuthal Fourier harmonics $m_{max}$ that enters
 to the force. To prevent coupling of modes through aliases,
 $2(m_{max}+1)$
particles are needed on each ring.
Quiet starts are also possible for warm stellar disks, but it
 is not practicable to suppress both
radial and azimuthal density variations at the same time
 (\cite{sell83,sell_ath86}).
Noise in the azimuthal forces must be suppressed since we are interested in 
non-axisymmetric instabilities. One then gives
 the particles on the ring identical
velocity components so that the initial orbits remain
 congruent.
\par
Choosing integrals of motion for each particle at random from the distribution function will result
in statistical fluctuation about the intended function. These can be eliminated by choosing
integrals for each particle in a deterministic manner such that their distribution is as close as possible
to the required form. For example, one can use
 energy and angular momentum as the independent
variables (\cite{sell_ath86});
 however, any other set of isolating integrals in which the distribution
 function can be expressed would work equally well.
\par
Since we are interested in making a quantitative and systematic 
comparison between the stability 
of bare disks obeying MOND, and Newtonian disks with dark halos we want to minimize the statistical
noise and employ a quiet start technique. As discussed above, one needs
  to use only a selected number of azimuthal Fourier components of 
$\Sigma$ in the 
force determination. In Newtonian dynamics this is justified for linear stability analysis
since the Poisson equation and the linearized collisionless Boltzmann
 equation do not couple
modes with different azimuthal frequencies. The 
MOND field equation can be linearized around the solution of the
 unperturbed axisymmetric
disk (as discussed in \cite{mgst}) and together with the linearized
 collisionless Boltzmann
equation have the property that  unstable modes with  different
 azimuthal Fourier components
are uncoupled. Instead of using the linearized MOND equation we use
 the full MOND equation,
but leave only the desired Fourier components in the surface density
 that is assigned to
the grid. In setting up the initial conditions we use the following
 procedure: We take the
numerical solution obtained for the distribution function and 
interpolate  it to a finer
grid. We then calculate the number of particles that should
 reside in each cell given the 
total  number of particles. This number is usually not an
 integer, we take the integer part
and distribute the particles uniformly in the cell. The remaining
 fraction is interpreted
as the probability for an additional particle to reside in this cell.
 We then draw cells
at random according to their relative probabilities and place at most
 one additional
particle in a cell. At the end of this stage we have a list of the
 $ (r_{min},r_{max})$
coordinates of the particles. We now need to assign the phase-space
 coordinates for each
particle i.e. $r,\theta , v_r,v_{\theta}$. (Here we use a polar grid in the
mid-plane as an auxiliary for computing the discretized density distribution
that serves as input for the MOND field equation.)
 We draw randomly the radial
 position of the particle taking
the probability density of finding the particle at radius $r$ as being
 proportional
to $ v_r^{-1}$. If we decide to use the Fourier filtering we draw at
 random the  angle
$\theta$ and place $2(m_{max}+1)$ particles at angular spacings of
 $\pi(m_{max}+1)^{-1}$
adding a small random angular shift to each particle to seed the
 unstable modes. If we do not
use the Fourier filtering  we place the particle at a random angle
 $\theta$ requiring that
the surface density produced on the grid that is used by the potential
  solver will be as smooth
as possible.

\clearpage

\begin{table}
\centerline{\begin{tabular}{||c|c|c|c|c|c||} \hline
  m   &  Q   &time step& \multicolumn{2}{c|}{Growth rate} & halo mass   \\ 
      &      &scaling & MOND              & Newt+DM      & at R=1      \\ \hline
0.005 & 2.55 &  1     &                   &              &              \\
0.01  & 2.5  &  0.84  &    0.4            &              &              \\
0.02  & 2.4  &  0.7   &    0.43           &              &              \\
0.04  & 2.25 &  0.58  &    0.46           &  0.09        &   0.18       \\
0.08  & 2.0  &  0.48  &    0.51           &  0.36        &   0.23       \\
0.16  & 1.79 &  0.39  &    0.62           &  0.53        &   0.28       \\
0.32  & 1.62 &  0.3   &    0.8            &  0.8         &   0.31       \\
0.64  & 1.53 &  0.22  &    0.94           &  0.94        &   0.31       \\
1.28  & 1.5  &  0.16  &    1.0            &  0.97        &   0.27       \\ \hline
\end{tabular}}
\caption{The growth rate, in units of dynamical frequency,
for the $m=2$ mode, and model parameters for the different mass models.} 
\label{table1}
\end{table}


\clearpage
\newpage

\begin{thebibliography}{}
\bibitem[Bai and Brandt
 1987]{achi-local-ref1987}{ Bai, D. and Brandt, A.
 1987,  SIAM J. Sci. Stat. Comput., 8, No.2,}
\bibitem[Bekenstein and Milgrom (1984)]{mgbk}{ Bekenstein, J. and
  Milgrom, M.
 1984  ApJ, 286, 7}
\bibitem[Binney and Tremaine 1987]{bt}{Binney, J. and  Tremaine, S.
1987, {\it Galactic Dynamics}, Princeton University Press}
\bibitem[Brada and Milgrom 1995]{bra95}{Brada, R. and Milgrom, M.
1995 MNRAS, 276, 453}
\bibitem[Brandt 1977]{achi1977} {Brandt, A. 1977, Math. Comp., 31, pp. 333,
 ICASE Report 76-27, Hampton, VA}
\bibitem[Brandt 1984]{achi-mgguide}{ Brandt, A. 1994, Multigrid Techniques:
 1984 Guide,with Applications to Fluid Dynamics, monograph. Weizmann
 Institute of Science, Rehovot, Israel}
\bibitem[Brandt 1991]{achi-lattice91}{Brandt, A. 1992, Nuclear Physics B
 (Proc. Suppl.), 26,  137, North-Holland}
\bibitem[Christodoulou (1991)] {chst}{Christodoulou, D.M.
 1991, ApJ, 372, 471}
\bibitem[Earn and Sellwod (1995)]{earn_sell_1995}{Earn, D.J.D. and
Sellwood, J.A. 1995, ApJ, 451, 533}
\bibitem[Goldreich and Lynden-Bell 1965]{gold-lynd}{Goldreich, P. and
Lynden-Bell, D. 1965, MNRAS, 130, 125}
\bibitem[Griv and Zhytnikov (1995)]{griv95}{Griv, E. and
Zhytnikov, V.V. 1995, Astrophys. Sp. Sci., 226, 51} 
\bibitem[Hockney and Eastwood (1988)]{hoea}{Hockney, R.W. and Eastwood, J.W.
1988,  Computer simulation using particles, Adam-Hilger}
\bibitem[Hunter 1992]{hunt}{ Hunter, C. 1992, in Dermott S. F., 
Hunter J. H., Wilson R. E., eds., in Astrophysical disks. Ann. N.Y.
 Acad. Sci. p. 22}
\bibitem[Julian and Toomre 1966]{jul-toom}{Julian, W.H. and Toomre, A.
 1966, ApJ, 146, 810}
\bibitem[Kalnajs (1972)]{kaln72}{Kalnajs, A.J.  1972, ApJ, 175, 63}
\bibitem[Kalnajs (1977)]{kaln77}{Kalnajs, A.J.  1977, ApJ, 212, 637}
\bibitem[Kalnajs (1978)]{kalnajs1978}{Kalnajs, A.J.  1978, in IAU Symp. 77,
  Structure and Properties of Nearby
Galaxies, ed. E.M. Berkhuisjen and R. Wielebinski (Dordrecht: Reidel),113} 
\bibitem[McGaugh 1996]{mcgaugh96}{ McGaugh, S.S. 1996, MNRAS, 280, 337}
\bibitem[Milgrom 1986]{mgfe}{Milgrom, M. 1986, ApJ, 302, 617}
\bibitem[Milgrom 1989]{mgst}{Milgrom, M. 1989, ApJ, 338, 121}
\bibitem[Milgrom 1994]{mgMV}{Milgrom, M. 1994, ApJ, 429, 540}
\bibitem[Ostriker and Peebles (1973)]{ost_peeb_crit}{Ostriker, J.P. and
 Peebles, P.J.E.  1973, ApJ, 186, 467}
\bibitem[Sawamura 1983]{saw}{Sawamura, M. 1988, PASJ, 40, 27}
\bibitem[Sellwood 1983]{sell83}{Sellwood, J.A. 1983, J. Comp. Phys., 50, 337}
\bibitem[Sellwood 1986]{sell86}{Sellwood, J.A. 1986, MNRAS, 221, 213}
\bibitem[Sellwood 1987]{sell_rev87}{Sellwood, J.A. 1987, ARAA, 25, 151}
\bibitem[Sellwood and Athanassoula 1986]{sell_ath86}{Sellwood, J.A., and
 Athanassoula, E. 1986, MNRAS, 221, 195}
\bibitem[Sellwood and Wilkinson (1993)]{sell-wilk}{Sellwood, J.A.,  and
Wilkinson, A. 1993,  Rep. Prog. Phys., 56, 173}
\bibitem[Toomre (1981)]{toom_swing}{Toomre, A. 1981,
 in  The Structure and Evolution of Normal Galaxies, ed.
 S.M. Fall and D. Lynden-Bell (Cambridge: Cambridge University Press.), p 111} 

\end{thebibliography}
\end{document}